\documentclass[iop]{emulateapj}
\usepackage{epsfig}
\usepackage{apjfonts}
\usepackage{aas_macros}

\begin{document}

\title{GRB 080503 late afterglow re-brightening: signature of a magnetar powered merger-nova}
\author{He Gao\altaffilmark{1,2,3,4}, Xuan Ding\altaffilmark{1}, Xue-Feng Wu\altaffilmark{1,5}, Zi-Gao Dai \altaffilmark{6,7} and Bing Zhang\altaffilmark{8}}

\altaffiltext{1}{Purple Mountain
Observatory, Chinese Academy of Sciences, Nanjing 210008, China}
\altaffiltext{2}{Department of Astronomy and Astrophysics, Pennsylvania State University, 525 Davey
Laboratory, University Park, PA 16802}
\altaffiltext{3}{Department of Physics, Pennsylvania State University, 525 Davey Laboratory, University
Park, PA 16802}
\altaffiltext{4}{Center for Particle and Gravitational Astrophysics, Institute for Gravitation and the Cosmos,
Pennsylvania State University, 525 Davey Laboratory, University Park, PA 16802}
\altaffiltext{5}{Joint Center for Nuclear Particle Physics and Cosmology, Purple Mountain Observatory -
Nanjing
University, Nanjing 210008, China}
\altaffiltext{6}{School of Astronomy and Space Science, Nanjing University, Nanjing 2100093, China}
\altaffiltext{7}{Key Laboratory of Modern Astronomy and Astrophysics (Nanjing University), Ministry of
Education, China}
\altaffiltext{8}{Department of Physics and Astronomy, University of Nevada Las Vegas, NV 89154, USA}
\email{hug18@psu.edu; xfwu@pmo.ac.cn;dzg@nju.edu.cn;zhang@physics.unlv.edu}

\begin{abstract}
GRB 080503 is a short gamma-ray burst (GRB) detected by \emph{Swift} and has been classified as a compact-star-merger-origin GRB. The soft extended emission and the simultaneous late re-brightening in both the X-ray and optical afterglow lightcurves raise interesting questions regarding its physical origin. We show that the broad-band data of GRB 080503 can be well explained within the framework of the double neutron star merger model, provided that the merger remnant is a rapidly-rotating massive neutron star with an extremely high magnetic field (i.e. a millisecond magnetar). We show that the late optical re-brightening is consistent with the emission from a magnetar-powered ``merger-nova''. This adds one more case to the growing sample of merger-novae associated with short GRBs. The soft extended emission and the late X-ray excess emission are well connected through a magnetar dipole spin-down luminosity evolution function, suggesting that direct magnetic dissipation is the mechanism to produce these X-rays. The X-ray emission initially leaks from a hole in the merger ejecta pierced by the short GRB jet. The hole subsequently closes after the magnetar spins down and the magnetic pressure drops below ram pressure. The X-ray photons are then trapped behind the mergernova ejecta until the ejecta becomes optically thin at a later time. This explains the essentially simultaneous re-brightening in both the optical and X-ray lightcurves. Within this model, future gravitational wave sources could be associated with a bright X-ray counterpart along with the mergernova, even if the short GRB jet beams away from Earth.
\end{abstract}


\section {Introduction}

The frequency range of the next generation gravitational-wave (GW) detectors, such as Advanced LIGO
\citep{ligo}, Advanced VIRGO \citep{virgo} and KAGRA \citep{kagra} interferometers, is designed to
uncover the final inspiral and merger of compact object binaries (NS-NS, NS-BH, BH-BH systems). Due to
the faint nature of GW signals, an associated electromagnetic (EM) emission signal coinciding with a GW
signal in both trigger time and direction could play a crucial role for confirming the astrophysical origin of
the GW signals and studying the astrophysical origin of the GW sources (e.g. host galaxy, distance, etc).

Short-duration $\gamma$-ray bursts (SGRBs) have long been proposed to originate from mergers of
compact object binaries \citep{paczynski86,eichler89,paczynski91,narayan92}. If so, SGRBs may provide
the brightest EM counterpart associated with events detected by those upcoming interferometers.
However, observations of SGRBs suggest that at least some of them are collimated into a small opening
angle \citep{burrows06,depasquale10}, so that most GW signals would not be detected together with
SGRBs \cite[e.g.][]{metzger12}. Lately, additional EM signatures of the compact binary mergers
(especially for NS-NS system) becomes a topic of growing interest \cite[][for a review]{berger14}.

Numerical simulations show that a mildly isotropic, sub-relativistic outflow could be ejected during the
merger of binary neutron stars, including the tidal tail matter during the merger and the matter from the
accretion disk \cite[e.g.][]{rezzolla11,rosswog13,bauswein13,hotokezaka13}. The typical mass and speed
of the ejecta are in the range of $10^{-4}-10^{-2}~{\rm M}_{\odot}$ and $0.1-0.3~c$, respectively
\citep{hotokezaka13}. Recently, several interesting EM signatures from the ejecta have been well studied, whose brightness are essentially determined by the properties of the left over remnant from the merger.

Usually, the merger product is assumed to be either a black hole or a temporal hyper-massive neutron
star which survives 10-100 ms before
collapsing into the black hole \citep[e.g.][]{rosswog03,aloy05,shibata05,rezzolla11,rosswog13}. In this
case, an optical/infrared transient is expected to be powered by radioactive decay from r-process
radioactive material \citep{lipaczynski98,kulkarni05,metzger10,barnes13}, henceforth we call it
a r-process-powered merger-nova\footnote{It is named as ``macro-nova" by \cite{kulkarni05} due to its
sub-supernova luminosity, or ``kilo-nova" by \cite{metzger10} due to its luminosity being roughly
$\sim10^3$ times of the nova luminosity. }. Besides this thermal emission, a long-lasting
radio emission is also expected from the interaction between the ejecta and the ambient medium,
although it is normally too weak to be detected \citep{nakar11,metzger12,piran13}. Such transients
are more isotropic than SGRBs. Depending on the direction of our line of sight, these transients could be
detected alone or to be accompanied by SGRBs,
provided that their luminosities are large enough \citep{metzger12}.
After several years of search \citep{bloom06,perley09,kocevski10}, an r-process-powerd merger-nova
was finally claimed to be detected in the infrared band with the Hubble Space Telescope (HST) for GRB 130603B
\citep{tanvir13,berger13}. More recently, \cite{yang15} re-examined the late afterglow data of GRB 060614 observed with HST, and found a significant F814W-band excess at $t\sim13.6$ days after the burst. They claimed that it is very likely another candidate of r-process-powered merger-nova. For both cases, the mergernova interpretation was based on one single data point.

Alternatively, it has long been proposed that the postmerger product could be a stable or super-massive
millisecond magneter, if the equation of state of nuclear matter is stiff enough and the total mass of the two
neutron stars is small enough \citep{dai06,fanxu06,gaofan06,zhang13,giacomazzo13}. Evidence of a
magnetar following some SGRBs has been collected in the Swift data, including the extended emission
\citep{norris06,metzger08}, X-ray flares \citep{barthelmy05b,campana06} and more importantly,
``internal plateaus" with rapid decay at the end of the plateaus \citep{rowlinson10,rowlinson13,lv15}.
Nonetheless, available observations (e.g., the lower limit of the maximum mass of Galactic NSs and the
total mass distribution of Galactic NS-NS binaries) and numerical simulations allow the existence of
the post-merger massive NS remnant
\cite[][and reference therein]{zhang13,metzger14}. Compared with the black
hole merger remnants, the main consequences of magnetar merger remnants include the following:
\begin{itemize}
\item The spin-down of the NS remnant supplies an additional energy source to the system;
\item The strong neutrino-driven wind from the NS provides additional mass outflow to the system.
\end{itemize}
In this case,  the detectable EM signatures from the system become much richer. Besides the putative short GRB signature, the following EM signals may be expected. First, the magnetar would eject a near-isotropic Poynting-flux-dominated outflow, the dissipation of which could power a bright early X-ray afterglow \citep{zhang13}. Second, the thermal emission from the ejecta could be significantly enhanced due to additional heating from magnetar wind \citep{yu13, metzger14}. This power could exceed the r-process power, so that we may call the corresponding transient as a ``magnetar-powered merger-nova''. Finally, the magnetar-power would energize and accelerate the ejecta to a mildly or even moderately relativistic speed, and the interaction between the ejecta and the ambient medium could produce a strong external shock that gives rise to bright broad-band emission (i.e. the double neutron star (DNS) merger afterglow model, \citealt{gao13GWB}). Some recently discovered transients could be interpreted within such a scenario, lending support to a post-merger magnetar remnant.
For instance, the Palomar Transient Factory (PTF) team recently reported the discovery of a rapidly fading optical transient source, PTF11agg. Lacking a high-energy counterpart, it has been proposed to be a good candidate for the DNS merger afterglow emission \citep{wu14}. Moreover, considering its broad-band data, GRB 130603B and its claimed ``kilonova'' can be interpreted within the framwork of a magnetar-powered DNS merger remnant given that the magnetar underwent significant energy lost through GW radiation
\citep{fan13c,metzger14}.

Similar to GRB 130603B, GRB 080503 is a SGRB with bright extended emission. Based on its negligible
spectral lag of prompt emission and extremely faint afterglow, GRB 080503 has been classified as a
compact-star-merger-origin GRB\footnote{The physical category of a GRB may not always be straightforwardly inferred based on the duration information, and multi-band observational criteria are needed (Zhang et al. 2009).} \citep{perley09} .  The most peculiar feature in GRB 080503 is that after the
prompt emission (began with a short spike and followed by extended emission) and the early steep
decay afterglow phase, it didn't immediately enter into the regular afterglow phase. Being signal-less
for about one day, it presented a surprising re-brightening in both the optical and X-ray bands. In the optical,
it remained bright for nearly five days. Within the post-merger remnant is a black hole, the scenario has been
investigated for GRB 080503. A ``r-process-powered merger-nova" model can marginally explain the
optical data, but the X-ray data could not be interpreted \citep{perley09,hascoet12}.
In this work, we make a comprehensive
analysis on the multi-band observations of GRB 080503, and suggest that the magnetar merger remnant scenario
can well interpret the entire data set, making a solid case to connect the late optical excess of GRB 080503
with a magnetar-powered merger-nova. We note that the idea that GRB 080503 is a good candidate for a
magnetar-powered transient has been qualitatively proposed by \cite{metzger14}.

\section{Obervational features of GRB\,080503}
GRB\,080503 was detected by Burst Alert Telescope (BAT) on abroad \emph{Swift} satellite at
12:26:13 on 2008 May 3 (see observational details in \cite{perley09}). Its prompt emission (in the 15-150
keV bandpass) contains a short bright initial spike with a duration of 0.32 $\pm$ 0.07 s,  followed by a
soft extended emission lasting for 232 s. The peak flux of the initial spike (measured in a 484~ms time
window) and the fluence of the extended emission (measured from 5~s to 140~s after the BAT trigger) are
$(1.2 \pm 0.2) \times 10^{-7}$ erg cm$^{-2}$ s$^{-1}$ and $(1.86 \pm 0.14) \times 10^{-6}$ erg cm
$^{-2}$, respectively. Although the fluence ratio between the extended emission and the spike is as large
as 30 in $15-150~ \rm {keV}$, \cite{perley09} further analyzed its features of hardness ratio and spectral
lag in detail, and found that this burst is still more reminiscent of a compact-star-merger-origin GRB.

After the extended emission phase, the X-ray light curve decays rapidly ($\alpha = 2$--4, where $F_{\nu}
\propto t^{-\alpha}$) until below the XRT detection threshold, and kept undetectable for about 1 day (as
shown in Figure \ref{fig:fit}). Then the X-ray flux rebrightened to the level of $10^{-3}~\mu {\rm Jy}$
around $10^{5}~{\rm s}$ after the BAT trigger. 20 days later, the Chandra X-Ray Observatory ACIS-S was
again employed to conduct imaging on the relevant position, but no source was detected.

In the optical band, many facilities were employed to search for afterglow signals on the first night after the
trigger, such as Swift UV-Optical Telescope (UVOT), Keck-I telescope and Gemini-North telescope, only a
single Gemini g band detection was obtained at 0.05 day. However, on the second night after the trigger,
the afterglow surprisingly rise above the detection threshold to the level of $10^{-1}~\mu {\rm Jy}$ and
kept bright for nearly five days. Later on, the localization region was observed with Hubble Space
Telescope in two epochs on 2008 May 12 and July 29. Although only upper limits were achieved, the
results infer a rapid decay feature for the late optical excess component.

During the observation, many attempts to measure the redshift of GRB 080503 were operated, even with
Hubble Space Telescope. Unfortunately, only an upper limit, $z<4$, was achieved.

\section{Model description}

\subsection{General picture}

If the equation of state of nuclear matter is stiff enough, the central product for a binary neutron star
merger could be a stable or a supra-massive NS rather than a black hole. This newborn massive NS would be
rotating with a rotation period in the order of milliseconds (close to the centrifugal break-up limit), and may
also contain a strong magnetic field $B \gtrsim 10^{14}\,{\rm G}$ similar to ``magnetars" \cite[]
[and reference therein]{zhang13,metzger14}. The millisecond magnetar is surround by a sub-relativistic
($v_{\rm ej}\sim 0.1-0.3c$) ejecta with mass $\sim (10^{-4}-10^{-2}) {\rm M_\odot}$
\citep{hotokezaka13}. Considering a variety of the origins for the ejecta materials, a spherical symmetry
could be reasonably assumed for the ejecta \citep{metzger14}.

Shortly after the formation, the magnetar would be surrounded by a centrifugally supported accretion disc
\citep{metzger08,dessart09,lee09,fernandez13}, launching a short-lived ($\lesssim~{\rm s}$)
collimated jet \citep{zhangd08,zhangd09,zhangd10}. The jet could easily punch through the ejecta shell and then power the prompt short spike
emission and the broad band GRB afterglow emission \citep{metzger08}. After the whole jet passing
through, it is possible that the gap remains open as the Poynting-flux-dominated magnetar wind continuously
penetrates through the hole. Due to the dynamical motion of the ejecta, the ejecta materials tend to quench
the outflow by closing the gap. During the early spindown phase when the Poynting-flux luminosity is
essentially a constant, the ram pressure of the ejecta may be balanced by the magnetic pressure of the
outflow. After the characteristic spindown time scale $t_{\rm sd}$, the magnetic pressure drops quickly,
so that the gap is closed in a time scale of $t_{\rm close}$. The total duration when the magnetar wind
leaks from the ejecta and make bright X-ray emission due to internal dissipation (e.g. through an
internal-collision-induced magnetic reconnection and turbulence (ICMART) process, \citealt{zhangyan11})
is the duration of the extended emission, i.e. $t_{\rm ee} = t_{\rm sd} + t_{\rm close}$. According to a more
detailed estimate (see Appendix for details), this duration can be consistent with the observed duration of
extended emission given reasonable parameters.
After $t_{\rm ee}$, the magnetar wind is stifled behind the ejecta, so that soft extended emission stops and the high-latitute emission in X-ray band shows up, which gives a rapidly dropping X-ray tail \citep{kumar00,zhangbb07,zhangbb09}. It is similar to the situation when the central engine is shut down.

Trapped by the ejecta materials, the magnetar continuously spin down, and when the magnetar wind encounters
the ejecta, a significant fraction of the wind
energy (parameterized as $\xi$) could be deposited into the ejecta, either via direct energy injection by a
Poynting flux \citep{bucciantini12}, or due to heating from the bottom by the photons generated in a
dissipating magnetar wind via forced reconnection (if $R<R_{\rm dis}$) or self-dissipation (if $R>R_{\rm
dis}$) \citep{zhang13}. Such continuous energy injection not only heats the ejecta material to power the merger-nova, \citep{yu13,metzger14}, but also accelerates the ejecta to a mildly or moderately relativistic speed, giving rise to strong afterglow emission by driving a strong forward shock into the ambient medium \citep{gao13GWB}. Note that the remaining fraction of the wind energy ($1-\xi$) would be stored in the trapped dissipation photons and eventually diffuse out with a deducing factor $e^{-\tau}$, where $\tau$ is the optical depth of the ejecta.

\begin{figure*}
\centering
\includegraphics[width=0.6\textwidth, angle =90]{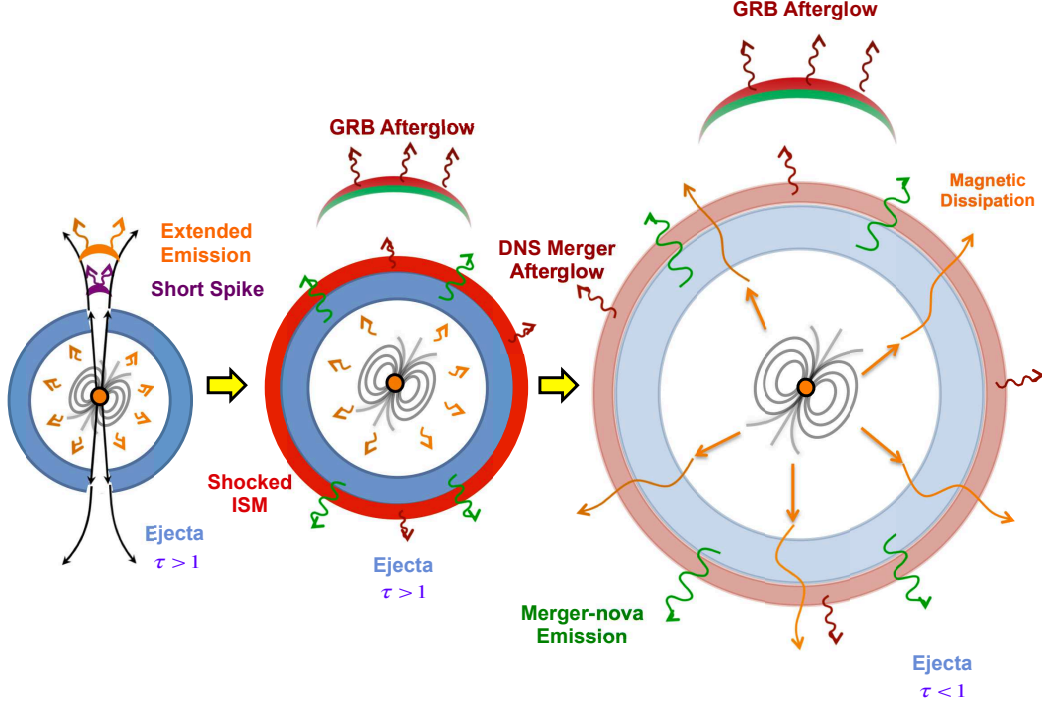}
\caption{A cartoon picture of model, illustrating various emission sites and emission components at different stages, from a NS-NS merger event that results in a stable millisecond magnetar remnant. In the early stage, a short-lived jet was launched, punching through the ejecta shell and giving rise to the prompt short spike emission. Following the jet, the magnetar wind leaks out through the opening gap, dissipates at a larger radius, and powers the extended emission. In the intermediate stage, the magnetar spins down and the ram pressure overcomes the magnetic pressure, so that the gap is closed due to the hydrodynamical motion of the ejecta. The magnetar wind is trapped behind the ejecta, which heats and accelerates the ejecta, powering the merger-nova emission and the DNS merger afterglow emission. In the mean time, the initial jet energy would drive GRB afterglow emission. In the late stage, the ejecta becomes transparent, the magnetar wind still dissipates its energy and radiate X-ray photons, which freely escape the remnant and give rise to the rebrightening in the X-ray lightcurve.}
\label{fig:cartoon}
\end{figure*}

In summary, there are four emission sites and several emission components involved in this model (as shown in Figure \ref{fig:cartoon}): i) the jet component that powers the short spike in prompt emission and the GRB afterglow emission; ii) the early magnetar wind component that powers the soft extended emission and the high latitude tail emission; iii) the magnetar-powered merger-nova emission acomponent, and the corresponding DNS merger afterglow emission; iv) the late
magnetar-wind-powered X-ray component when the ejecta becomes transparent. In the following, we describe the details for calculating these main emission components.

\subsection{Magnetar wind dissipation}

Considering a millisecond magnetar with an initial spin period $P_i$ and a dipolar magnetic field
of strength $B$, its total rotational energy reads $E_{\rm{rot}}=(1/2)I
\Omega_{0}^{2} \simeq 2\times 10^{52} I_{45} P_{i,-3}^{-2} ~{\rm
erg}$ (with $I_{45} \sim 1.5$ for a massive neutron star). The spin-down luminosity of the magnetar as a
function of time could be expressed as
\begin{eqnarray}
L_{\rm sd}=L_{\rm sd,i}\left(1+{t\over t_{\rm
sd}}\right)^{-2}
\end{eqnarray}
where
\begin{eqnarray}
L_{\rm sd,i}=10^{47}~R_{s,6}^6B_{14}^{2}P_{i,-3}^{-4}\rm~erg~s^{-1}
\end{eqnarray}
is the initial spin-down luminosity, and
\begin{eqnarray}
t_{\rm sd}=2\times10^{5}~R_{s,6}^{-6}B_{14}^{-2}P_{i,-3}^{2}~\rm s
\end{eqnarray}
is the spin-down timescale. Hereafter the convention $Q_x=Q/10^x$ is adopted in cgs units.

The spin-down luminosity is essentially carried by a nearly isotropic Poynting-flux-dominated outflow. In the free zone (e.g. in the direction of the cavity drilled by the jet, or the intrinsically open regions in the ejecta), the magnetar wind would leak out from the ejecta, and undergo strong self-dissipation beyond $R_{\rm dis} > R_{\rm ej}$, giving rise to extended emission (along the jet direction) or a bright X-ray afterglow emission \cite[off-axis direction,][]{zhang13}. In the confined wind zone, the magnetar wind is expanding into the ejecta, and the magnetic energy may be rapidly discharged via forced reconnection (if $R<R_{\rm dis}$) or self-dissipation (if $R>R_{\rm
dis}$). The trapped dissipation photons would eventually show up when the ejecta becomes optically thin.

As a rough estimation, one can assume an efficiency factor $\eta_{\nu}$ to convert the spin-down
luminosity to the observed luminosity at frequency $\nu$, so that one has
\begin{eqnarray}
F_{\nu} & \sim & \frac{\eta_{\nu} L_{\rm sd}}{4 \pi \nu D_L^2},
\end{eqnarray}
In this work, we take $\eta_{\nu}=0.3$ for both the extended emission and
the late dissipated emission.

\subsection{Magnetar powered merger-nova}

Suppose the magnetar is surrounded by a quasi-spherical ejecta shell with mass $M_{\rm ej}$ and initial
speed $v_i$. A generic model for the dynamics and emission properties of the ejecta could be briefly summarized as follows \cite[e.g.][]{yu13}.

Considering the energy injection from the magnetar and the energy dissipation through sweeping up the ambient medium, the total ``effective kinetic energy'' (total energy minus rest-mass energy) of the system can be expressed as
\begin{eqnarray}
E=(\Gamma-1)M_{\rm ej}c^2+\Gamma E'_{\rm int}+(\Gamma^2-1)M_{\rm sw}c^2,
\label{eq:Etot}
\end{eqnarray}
where $\Gamma$ is the bulk Lorentz factor of the ejecta, $E'_{\rm int}$ is
the internal energy measured in the comoving rest frame, $M_{\rm sw}=\frac{4\pi}{3}R^3nm_p$ is the
swept mass from the interstellar medium (with density $n$) and $R$ is the radius of the ejecta. The dynamical evolution of the ejecta can be
determined by
\begin{eqnarray}
{d\Gamma\over dt}={{dE\over dt}-\Gamma {\cal D}\left({dE'_{\rm int}\over
dt'}\right)-(\Gamma^2-1)c^2\left({dM_{\rm sw}\over dt}\right)\over
M_{\rm ej}c^2+E'_{\rm int}+2\Gamma M_{\rm sw}c^2}
\label{eq:Gt}
\end{eqnarray}
where ${\cal D}=1/[\Gamma(1-\beta)]$ is the Doppler factor with $\beta=\sqrt{1-\Gamma^{-2}}$. The
comoving time $dt'$ and luminosity $L'$ can be connected with the observer's time and luminosity by
$dt'={\cal D}dt$ and $L'={\cal D}^{-2}L$, respectively.

With energy conservation, we have
\begin{eqnarray}
{dE\over dt}=\xi L_{\rm sd}+{\cal D}^{2}L'_{\rm ra}-{\cal D}^{2}L'_{\rm e}.
\label{eq:Et}
\end{eqnarray}
The radioactive power $L'_{\rm ra}$ reads
\begin{eqnarray}
L'_{\rm ra}=4\times10^{49}M_{\rm ej,-2}\left[{1\over2}-{1\over\pi}\arctan \left({t'-t'_0\over
t'_\sigma}\right)\right]^{1.3}~\rm erg~s^{-1},
\label{eq:Lrap}
\end{eqnarray}
with $t'_0 \sim 1.3$ s and $t'_\sigma \sim 0.11$ s \citep{korobkin12}. The radiated bolometric
luminosity $L'_e$ reads \footnote{The energy loss due to shock emission is ignored here, as is usually done
in GRB afterglow modeling.}
\begin{eqnarray}
L'_e=\left\{
\begin{array}{l l}
  {E'_{\rm int}c\over \tau R/\Gamma}, & \tau>1, \\
  {E'_{\rm int}c\over R/\Gamma}, &\tau<1,\\ \end{array} \right.\
  \label{eq:Lep}
\end{eqnarray}
where $\tau=\kappa (M_{\rm ej}/V')(R/\Gamma)$ is the optical depth of the ejecta with $\kappa$ being
the opacity \citep{kasen10,kotera13}.

The variation of the internal energy in the comoving frame can be expressed by \cite[e.g.][]{kasen10}
\begin{eqnarray}
{dE'_{\rm int}\over dt'}=\xi {\cal D}^{-2}L_{\rm sd}+ L'_{\rm ra} -L'_{\rm e}
-\mathcal P'{dV'\over dt'},
\label{eq:Ep}
\end{eqnarray}
where the radiation dominated pressure can be estimated as $\mathcal P'=E'_{\rm int}/3V'$ and the
comoving volume evolution can be fully addressed by
\begin{eqnarray}
{dV'\over dt'}=4\pi R^2\beta c,
\label{eq:Vp}
\end{eqnarray}
together with
\begin{eqnarray}
{dR\over dt}={\beta c\over (1-\beta)}.
\label{eq:rt}
\end{eqnarray}

A full dynamical description of the system as well as the bolometric radiation luminosity can be easily
obtained by solving above differential equations. Assuming a blackbody spectrum for the
thermal emission of the mergernova, for a certain observational frequency $\nu$, the observed flux can be calculated as
\begin{eqnarray}
F_{\nu}={1\over4\pi D_L^2\max(\tau,1)}{8\pi^2  {\cal D}^2R^2\over
h^3c^2\nu}{(h\nu/{\cal D})^4\over \exp(h\nu/{\cal D}kT')-1},
\end{eqnarray}
where $h$ is the Planck constant.

\subsection{GRB afterglow emission}

The interaction between the initial launched jet and the ambient medium could generate a strong
external shock, where particles are believed to be accelerated, giving rise to broad-band synchrotron
radiation \cite[][for a review]{gao13review}. The total effective kinetic energy of the jet and the medium can be expressed as
\begin{eqnarray}
E=(\Gamma-1)M_{\rm jet}c^2+(\Gamma^2-1)M_{\rm sw}c^2,
\end{eqnarray}
where $M_{\rm sw}=2\pi(1-\cos\theta)/3R^3nm_p$ with $\theta$ being the half opening angle of the
jet. The energy conservation law gives
\begin{eqnarray}
{d\Gamma\over dt}={-(\Gamma^2-1)\left({dM_{\rm sw}\over dt}\right)\over
M_{\rm jet}+2\Gamma M_{\rm sw}}.
\label{eq:dyngrb}
\end{eqnarray}
where the energy loss due to shock emission is ignored.

In the co-moving frame, synchrotron radiation power at frequency $\nu '$ from electrons is given by
(Rybicki \& Lightman 1979)
\begin{equation}
\label{eq:pnup}
P'_{\nu'} = \frac{\sqrt{3} q_e^3 B'}{m_{\rm e} c^2}
	    \int_{\gamma_{\rm e,m}}^{\gamma_{\rm e,M}}
	    \left( \frac{dN_{\rm e}'}{d\gamma_{\rm e}} \right)
	    F\left(\frac{\nu '}{\nu_{\rm cr}'} \right) d\gamma_{\rm e},
\end{equation}
where $q_e$ is electron charge,
$\nu_{\rm cr}' = 3 \gamma_{\rm e}^2 q_e B' / (4 \pi m_{\rm e} c)$ is the
characteristic frequency of an electron with Lorentz factor $\gamma_e$, $B'$ is the comoving magnetic
field strength and
\begin{equation}
\label{fx23}
F(x) = x \int_{x}^{+ \infty} K_{5/3}(k) dk,
\end{equation}
with $K_{5/3}(k)$ being the Bessel function.

The comoving magnetic field strength $B'$ could be estimated as
\begin{eqnarray}
B'=(8 \pi e_s\epsilon_B)^{1/2},
\end{eqnarray}
where $e_s$ is the energy density in the shocked region and $\epsilon_B$ is the fraction of the shock
energy density that goes into the magnetic field.

The distribution of the shock-accelerated electrons behind the blast wave is usually assumed to be a
power-law function of electron energy,
\begin{equation}
\frac{dN_{\rm e}'}{d\gamma_{\rm e}} \propto \gamma_{\rm e}^{-p},
\,\,\,\,\,\,(\gamma_{\rm e,m}\leq \gamma_{\rm e} \leq\gamma_{\rm e,M}),
\end{equation}
Assuming that a constant fraction $\epsilon_e$ of the shock energy is distributed to electrons, the
minimum injected electron Lorentz factor can be estimated as
\begin{eqnarray}
\gamma_{\rm e,m}=g(p)\epsilon_e(\Gamma -1) \frac {m_p} {m_e},
\end{eqnarray}
where the function $g(p)$ takes the form
\begin{eqnarray}
\label{gp} g(p) \simeq \left\{ \begin{array}{ll} \frac{p-2}{p-1}, & p>2;\\
\rm{ln}^{-1}(\gamma_{\rm e,M}/\gamma_{\rm e,m}), &
p=2. \\
\end{array} \right.
\end{eqnarray}
The maximum electron Lorentz factor $\gamma_{e,M}$ could be estimated by balancing the acceleration time scale and the dynamical time scale, i.e.
\begin{eqnarray}
\gamma_{\rm e,M}\sim \frac{\Gamma t q_e B}{\zeta m_p c},
\end{eqnarray}
where $\zeta\sim 1$ is a parameter that describes the details of acceleration.
 If the electron energy has a harder spectral index $1 < p < 2$, the minimum electron Lorentz factor
 would be derived as \citep{daicheng01,bhattacharya01}
\begin{eqnarray}
\gamma_{\rm e,m}=\left(\frac{2-p}{p-1}\frac{m_p}{m_e}\epsilon_e(\Gamma
-1)\gamma_{\rm e,M}^{p-2}\right)^{1/(p-1)}
\end{eqnarray}

With the dynamical description of the jet and these radiation equations, one can calculate the evolution of $P'_{\nu'}$. Assuming that this power is radiated isotropically, then the observed flux density at frequency $\nu={\cal D}\nu'$ can be calculated as
\begin{eqnarray}
F_{\nu}={{\cal D}^3\over4\pi D_L^2}P'_{\nu'},
\label{eq:fnuobs}
\end{eqnarray}
Note that in the following calculation, we neglect the sideway expansion of the jet \citep{rhoads99,sari99}, but consider the jet break with edge effect at later time when $\Gamma^{-1}>\theta$ \citep{panaitescu98}.

\subsection{DNS merger afterglow emission}

During the propagation of the ejecta, a strong external shock would also form upon interaction with the
ambient medium. With dynamical solution for the ejecta, and radiation equations \ref{eq:pnup} to \ref{eq:fnuobs}, one can easily  calculate the relevant broad-band DNS merger afterglow emissions \citep{gao13GWB}.

\section{Application to GRB 080503}

Considering the extremely deep limit for the host galaxy of GRB 080503, one plausible possibility is that
GRB 080503 is at a moderately high redshift in a very underluminous galaxy, e.g. $z\approx 1$,
comparable to the highest-z SGRBs \citep{perley09}. In the following, we adopt $z=1$,
and investigate the broad-band
data of GRB 080503 with the physical model proposed in the last section. We find that with all standard
parameter values, the broadband data of GRB 080503 could be well explained. The fitting results (data
from R. Hascoet et al. 2012) are presented in Figure \ref{fig:fit} and the adopted parameter values are
collected in Table 1. We briefly summarize the investigation results as follows:
\begin{figure}
\includegraphics[width=3.5in]{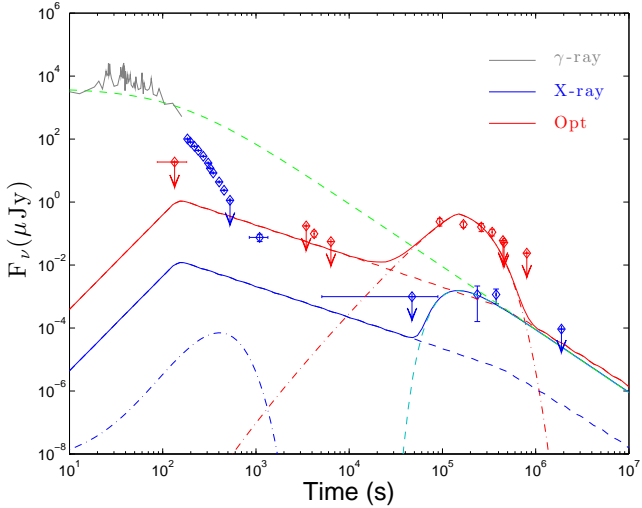}
\caption{Modeling results for the broad-band observations of GRB 080503. The data are taken from R. Hascoet et al. (2012), with blue denoting X-rays and red denoting optical. Blue and red dashed lines represent the GRB afterglow emission in the X-ray and optical bands, respectively; blue and red dotted dash lines represent the merger-nova emission in the X-ray and optical band, respectively; the green dashed line denotes the evolution function of the magnetar spin-down radiation luminosity; the light blue dashed line denotes the late magnetar wind dissipation emission. The blue and red solid lines are final fitting lines for the X-ray and optical data, respectively.}
\label{fig:fit}
\end{figure}
\begin{itemize}
\item The soft extended emission and the late X-ray excess peak could be well connected
with a magnetar spin-down luminosity evolution function, suggesting direct magnetic dissipation as the
same underlying origin for these two observed components.
\item In the X-ray band, the contribution from the merger-nova and early GRB afterglow emission are
outshone by the aforementioned direct magnetic dissipation component;
\item The early optical data can be explained by the GRB afterglow emission. The late optical data,
including the re-brightening phase and the rapid decay feature, can be well explained by the emission
from a magnetar-powered merger-nova.
\item Both the late-time optical and X-ray data peak around the same time when $\tau=1$, which
is consistent with the argument that the late magnetar dissipation photons can travel freely after the
ejecta becomes transparent. This powers the late X-ray excess;
\item For this particular event, the DNS merger afterglow emission is completely suppressed, since the ejecta mass
is relatively large so that the ejecta is only mildly relativistic, and since the medium density is small.
This emission component is not plotted in Fig.\ref{fig:fit}.
\end{itemize}

In the interpretation, we adopt the isotropic kinetic energy of the jet as $10^{51}~{\rm erg}$, which is based on the total emission energy of the short spike and assume a factor of $20\%$ for the $\gamma$-ray emission efficiency. The values for initial Lorentz factor ($\Gamma_0$) and half opening angle ($\theta$) of the jet are chosen as 200 and 0.1, the values of which barely affect the final fitting results. To achieve the faintness of the GRB afterglow emission, a relatively low value for ambient medium density ($n=0.001~{\rm cm^{-3}}$) is required, suggesting that GRB 080503 may have a large offset relative to the center of its host galaxy, which in turn explains the extremely deep limit on its host at the afterglow location (since the system may have been kicked out far away from a host galaxy). The  microphysics shock parameters (e.g., $\epsilon_e$, $\epsilon_B$, $\zeta$ and $p$) are all chosen as their commonly used values in GRB afterglow modeling \cite[][for a review]{kumarzhang14}. For the magnetar, a relatively large stellar radius $R_{\rm s,6}=1.2$ is adopted by considering a rapidly rotating supra-massive NS, and initial spin period $P_i$ is taken as $2~{\rm ms}$ by
considering a mild angular momentum loss via strong gravitational radiation \citep{fan13b}. The dipolar magnetic field of strength $B$ is adopt as $6\times 10^{15}~{\rm G}$, which is consistent with the suggested values by fitting the SGRBs X-ray plateau feature \citep{rowlinson13,lv14b}. For the ejecta, we take the standard values of mass ($M_{\rm ej}\sim10^{-3}{\rm M_{\odot}}$) and initial velocity ($v_i=0.2c$), and a relatively large value for the effective opacity $\kappa=10~{\rm cm^{2}~g^{-1}}$. The latter was suggested by recent works by considering the bound-bound, bound-free, and free-free transitions of ions \citep{kasen13,tanaka13}. If the opacity has a smaller value (e.g. because of the intense neutrino emission from the porto-magnetar, \cite{metzger14b}), the same data can be interpreted by increasing the mass of the ejecta. For example, an equally good fit can be reached with $M_{\rm ej}\sim10^{-2}{\rm M_{\odot}}$ for $\kappa=1~{\rm cm^{2}~g^{-1}}$. Finally, we assume that $30\%$ of the wind energy is deposited into the ejecta, which is a nominal value suggested from previous works \citep{zhangyan11,yu13}.

\begin{table}
\begin{center}{\scriptsize
\caption{Parameters for interpreting the broadband data of GRB\,080503, by assuming z=1.}
\begin{tabular}{cccccc} \hline\hline
 \multicolumn{6}{c}{Magnetar and ejecta parameters}\\
  \hline
 $B~({\rm G})$                    & $P_i~({\rm ms})$   &$R_s~({\rm cm})$ &$M_{\rm ej}~({\rm M_{\odot}})$                    &  $v_i/c$   & $\kappa~({\rm cm^{2}~g^{-1}})$     \\
 $6\times10^{15}$     &  $2$         &$1.2\times10^{6}$ & $3\times10^{-3}$     &  $0.2$         &$10$\\
   \hline
  \multicolumn{6}{c}{Jet and ambient medium parameters}\\
  \hline
 &$E~({\rm erg})$     &$\Gamma_0$               &  $n~(\rm{cm^{-3}})$& $\theta~({\rm rad})$ \\
 &$10^{51}$     &  $200$ &  $0.001$   & $0.1$       \\
  \hline
  \multicolumn{6}{c}{Other parameters}\\
  \hline
 $\epsilon_e$                    &  $\epsilon_B$&  $p$ &$\zeta$& $\xi$ & $\eta_\gamma$\\
 $0.01$     &  $0.001$&  $2.3$ &$1$  & $0.3$ &   $0.3$    \\
   \hline\hline
 \end{tabular}
 }
\end{center}
\end{table}

\section{Conclusion and Discussion}
\label{sec:discussion}

Double neutron star mergers could leave behind a millisecond magnetar rather than a black hole. In this
scenario, the spin-down of the magnetar provides an additional energy source in the merger remnant,
which generates much richer EM signatures from the merger remnant system than the black hole scenario.
In this work, we give a
comprehensive description for all the possible EM signals under the magnetar remnant scenario, invoking
several emission sites to account for several emission components, i.e. the initially launched jet to
produce the short spike in prompt emission; an external shock site for this jet component to account
for part of the observed optical afterglow emission component; an magnetar wind internal dissipation site that accounts for the early soft extended emission, the high latitude tail emission, as well as the late X-ray re-brightening emission when the ejecta becomes transparent; an isotropic ejecta site that generates a magnetar-powered merger-nova emission; and finally the site where the ejecta interacts with the medium and powers the DNS merger afterglow emission.

We presnt the detailed numerical methods to calculate these emission components and apply the model
to investigate the broadband observations of GRB 080503. We find that the magentar remnant scenario
could well interpret the multi-band data of GRB 080503, including the extended emission and its re-brightening features in both X-ray and optical bands. In our calculation, we adopt $z=1$ for GRB 080503, which could be a plausible assumption in view of both the extremely deep upper limit for the host galaxy flux and the observed redshift distribution of SGRBs. If our interpretation is correct, some important implications could be inferred:
\begin{itemize}
\item GRB 080503 is of a double neutron star merger origin;
\item The post-merger remnant of this event is a stable magnetar, with an effectively polar cap dipole magnetic field $6\times 10^{15}$ G and an initial period $2$ ms;
\item The late optical re-brightening is a magnetar-powered merger-nova. Since its emission is essentially isotropic, similar merger-novae are expected to be associated with NS-NS merger gravitational wave sources even without a short GRB association;
\item For this event, the ejected mass during the merger is estimated to be around $3\times 10^{-3}~M_{\odot}$.
\end{itemize}

To justify the assumption of $z=1$, we also tested other redshift values (either smaller or
larger than 1). We find that the fitting results are not sensitive to the redshift value, even though
some parameters may vary within reasonable ranges.

In this work, we assume that the magetar wind is highly magnetized, i.e., with a high $\sigma$ value. If, on the other hand, the wind contains a significant fraction of primary $e^{\pm}$ pairs, the magnetic wind may become leptonic-matter-dominated upon interaction with the ejecta, so that a strong reverse shock can be developed, which would predict additional radiation signatures \citep{dai04,wanglj13,wanglj15}. Moreover, \cite{metzger14} proposed that the large optical depth of  $e^{\pm}$ pairs inside the ejecta shell could also suppress the efficiency for depositing the wind energy into the ejecta, which essentially corresponds to a reduced value of $\xi$ in our model.

In our interpretation, we assume that the magnetar wind could leak out from the ejecta shell through the opening gap drilled by the initial jet, powering the extended emission. An alternative interpretation could be that the outflow from the magnetar wind itself may be collimated into a bipolar jet by its interaction with this ejecta \citep{bucciantini12} and then power the extended emission \citep{metzger14}. If this is the case, the real spin-down luminosity would be smaller than the extended emission luminosity due to the collimation effect, inferring a somewhat lower dipole field. However, such a collimation effect is only significant for a large ejecta mass (say $>10^{-2}~{\rm M_{\odot}}$), which should not affect the results in this work, since the preferred ejecta mass for the case of GRB 080503 is relatively small ($\sim 10^{-3}~{\rm M_{\odot}}$).

Finally, it is worth pointing out that our described physical picture for the EM signatures from a NS-NS merger with a stable or supra-massive millisecond magnetar remnant could be applied to other cases of short GRBs and also the cases when the jet direction beams away from Earth. A systematic study of extended emission and internal plateau emission from short GRBs \citep{lv15} revealed many plateaus followed by a rapid decay. It would be interesting to systematically apply the model to these GRBs to constrain the model parameters. In most cases, no X-ray rebrightening is observed, which suggests that the magnetar is likely supra-massive, and has collapsed into a black hole before the ejecta becomes transparent. In the future, off-axis X-ray transients may be discovered to be associated with gravitational wave events due to NS-NS mergers \citep{zhang13}. Applying our model to these events can give more detailed predictions to the brightness of these X-ray transients, which is valuable for searching for EM counterparts of GW signals in the Advanced LIGO/Virgo era.

\acknowledgments

This work is supported by the National Basic Research Program ('973' Program) of China (grants 2014CB845800 and 2013CB834900), the National Natural Science Foundation of China (grants nos. 11322328, 11033002 and 11433009). H.G. acknowledges support by NASA NNX 13AH50G. XFW is partially supported by the One-Hundred-Talents Program, the Youth Innovation Promotion Association, and the Strategic Priority Research Program ``The Emergence of Cosmological Structures'' (grant no. XDB09000000) of of the Chinese Academy of Sciences, and the Natural Science Foundation of Jiangsu Province (grant no. BK2012890).

\appendix

The initial jet launched during the early accretion phase that powers the short GRB may have drilled a bipolar cavity in the ejecta. The subsequent magnetar wind following the short GRB also penetrate through this cavity power the extended emission. During this phase, the ram pressure around the cavity due to the dynamical motion of the ejecta would be initially balanced by the transverse magnetic pressure in the magnetar wind, i.e.
\begin{eqnarray}
\frac{L_{\rm sd,i}\phi}{4\pi R^2 c}\sim \frac{M_{\rm ej}v^2}{4\pi R^2 \Delta},
\end{eqnarray}
where $c$ is the speed of light, $\phi$ is the transverse magnetic pressure fraction, $R$ is the radius of the ejecta, and $\Delta$ is the thickness of the ejecta shell. Under this condition, the corresponding fluid speed in the transverse direction due to dynamical motion of the ejecta can be estimated as
\begin{eqnarray}
v &=& \left(\frac{L_{\rm sd,i}\phi \Delta}{M_{\rm ej}c}\right)^{1/2}\nonumber\\
&\approx& 0.04c~R_{s,6}^{3}B_{15}P_{i,-3}^{-2}M_{\rm ej}^{-1/2}v_{i,10}^{1/2}\phi^{1/2},
\end{eqnarray}
where $\Delta = v_i \Delta t$, with $v_i \sim 10^{10}~{\rm cm~s^{-1}}$ and $\Delta t = 1$ s \citep{bucciantini12}.
When $t>t_{\rm sd}$, the magnetic pressure quickly drops, so that the cavity would be gradually closed in a timescale
\begin{eqnarray}
t_{\rm close}\approx\frac{\beta_{\rm sd}t_{\rm sd}c\theta}{v_s}~,
\end{eqnarray}
where $\theta$ is the jet opening angle, $\beta_{\rm sd}$ is the ejecta radial speed at $t_{\rm sd}$. From energy conservation, we obtain
\begin{eqnarray}
\beta_{\rm sd}=\min \left[1,\left(\frac{\xi E_{\rm rot}}{M_{\rm ej}c^2}\right)^{1/2}\right].
\end{eqnarray}
For the cases with $\beta_{\rm sd}$ not close to unity (such as the case for GRB 080503 with $\beta_{\rm sd} \sim 0.5$), we have
\begin{eqnarray}
t_{\rm close}\approx 1.5\times 10^4~{\rm s}~ R_{s,6}^{-9}B_{15}^{-3}P_{i,-3}^{3}\theta_{-1}v_{i,10}^{-1/2}\xi^{1/2}\phi^{-1/2}.
\end{eqnarray}
The total time scale for the extended emission can be estimated as
\begin{eqnarray}
t_{\rm ee}=t_{\rm sd}+t_{\rm close}.
\end{eqnarray}
With the parameters adopted to interpret the data of GRB 080503, we have $t_{\rm ee}\approx74.4+157.3(\phi/ 0.25)^{-1/2}~{\rm s}$, which is well consistent with the stopping time of extended emission (232 s), provided that the $\phi \sim 1/4$ of the magnetic pressure concentrates in the transverse direction.

\end{document}